\documentclass[aps,pra,reprint,groupedaddress]{revtex4-1}
\usepackage{graphicx}
\usepackage{amsmath}
\usepackage{float}
\usepackage{hyperref}% add hypertext capabilities
\usepackage[mathlines]{lineno}
\usepackage{epstopdf}

\begin{document}

%\preprint{}

\title{On decoherence induced by a spin chain: role of initial prepared states}


\author{L. T. Kenfack}
\email[]{leonel.kenfack@univ-dschang.org}
\author{M. R. T. Fokou}
\author{M. Tchoffo}
\author{M. E. Ateuafack}
\author{L. C. Fai}
%\thanks{}
\affiliation{Research Unit of Condensed Matter, Electronic and Signal Processing, Department of Physics, Dschang school of Sciences and
Technology, University of Dschang, PO Box: 67 Dschang, Cameroon.}


\date{\today}

\begin{abstract}
We study the decoherence process induced by a spin chain environment on a central spin consisting of R spins and we apply it on the dynamics of quantum correlations (QCs) of three interacting qubits. In order to see the impact of the initial prepared state of the spin chain environment on the decoherence process, we assume the spin chain environment prepared in two main ways, namely, either the ground state or the vacuum state in the momentum space. We develop a general heuristic analysis  when the spin chain environment in prepared in these states, in order to understand the decoherence process against the physical parameters. We show that the decoherence process is mainly determined by the choice of the initial prepared state, the number of spin of the chain, the coupling strength, the anisotropy parameter and the position from the quantum critical point. In fact, in the strong coupling regime, the decoherence process does not appear for the environment prepared in the vacuum state and it behaves oscillatory in the case of evolution from ground state. On the other hand, in the weak coupling regime and far from the quantum critical point, decoherence induced by the ground state is weaker than that of the vacuum state. Finally, we show that, QCs are completely shielded from decoherence in the case of evolution from the W state and obey the same dynamics as the decoherence factors for the GHZ state.  
\end{abstract}

%\keywords{}

\maketitle

\section{\label{INTRODUCTION}INTRODUCTION}
In quantum world, every superposition of legal states is a legal state but when a system interacts with its surrounding, all its  superposed states are not treated equally. Indeed, the result of this interaction is the single out of pointer states which remain stable during the decoherence process, what is known as einselection \cite{Zurek}. The decoherence is then defined by Zurek as the destruction of quantum coherence between preferred states associated with the observables monitored  by the environment \cite{Zurek}.
\\ \hspace*{0.5cm}However, decoherence appears as the major obstacle to the conservation of quantum correlations (QCs) in quantum information processing. Indeed, QCs are the degree of quantum liaison among several systems. QCs appeared in last decade of 20th century from the concept of entanglement introduced by Schrodinger in 1935 \cite{Horedecki et al.}. In addition, there are two kinds of QCs measures : firstly, the quantumless of correlation such as concurrence, negativity, entanglement of formation and entanglement witness. What have mainly been introduced in the last decade of 20th century by Wooters \cite{Horedecki et al.}. Secondly, the quantumness of correlation which measure the rest of QCs after a set of measurements \cite{Ollivier et al.}.
\\ \hspace*{0.5cm}So many recent works have investigated on decoherence caused by environmental XY spin chain. These works can be separated into two sets: the former investigating the effect of XY spin chain on non-interacting central spins system \cite{Yan et al.,Yang et al.,Guo et al. 2013}, and the latter, the effect of XY spin chain on interacting central spins system \cite{You et al.,Hu,Guo et al. 2009,Guo et al. 2011,Xie et al.}.
\\ \hspace*{0.5cm}Concerning decoherence process induced by XY spin chain on non-interacting central spin, it has been established that, when the spin chain environment is prepared in the ground state, the decay of QCs due to decoherence differs following the initial prepared state of the central system, namely, a system of two and three qubits, in particular, in the case of a three qubits system, it is showed that, the W state is more robust against decoherence than GHZ state  \cite{Yan et al.,Guo et al. 2013}. In addition, in multipartite system, when only one qubit interacts with its surrounding, there is a sudden transition from classical to quantum decoherence \cite{Yan et al.,Yang et al.}.
\\ \hspace*{0.5cm}Concerning the decoherence process induced by XY spin chain on interacting central spins, it has been established that, when the spin chain environment is prepared in the ground state and the two-qubit system in separated one, the decoherence process is best enhanced at the quantum critical point in the weak coupling case and can be reduced both by decreasing the anisotropy of the chain and increasing the Dzyaloshinskii-Moriya (DM) interaction \cite{You et al.}. In the case of evolution from the vacuum state the disentanglement process is mainly determined by the physical parameters and the initial states of the central spins system, but it can be broken by introducing the Hamiltonian of the central spin system \cite{Hu}. However, there is another initial prepared state of the spin chain in the literature, called thermal state \cite{Guo et al. 2009,Guo et al. 2011,Jafarpour et al.}.
\\ \hspace*{0.5cm}In this work, we consider a central spins system consisting of $ R $ spins coupled to another spin chain environment. Each of them will be assumed to be prepared in two ways. We generalize the heuristic analysis made in ref.\cite{Yuan et al.,Cheng et al.,Sun et al.,Hu} for the spin chain's decoherence factors, in order to answer the following questions : How do the initial prepared states of the spin chain affect the decoherence process? How do the initial prepared states of the system affect the dynamics of QCs under such a decoherence process? How can the inevitable process of degradation of QCs due to decoherence be reversed? We start in Sec.~\eqref{s2} by introducing and diagonalizing the model and then we calculate the reduced density matrix of $ R $ qubits and explicit expressions of decoherence factors. In Sec.~\eqref{s3}, we generalize the heuristic analysis of decoherence factors and also present numerical results for two of them. In Sec.~\eqref{s4}, we investigate the QCs for $ R=3 $ and then present analytical and numerical results. Finally, Sec.~\eqref{s5} is devoted to the conclusion.

\section{MODEL}\label{s2}

	We model the spin chain environment by the one dimensional XY model:
\begin{equation}\label{one}
H_{E}^{\lambda} = -\sum_{l=1}^{N}\left[ \begin{split}
&
\frac{1+\gamma}{2}\sigma_{l}^{x}\sigma_{l+1}^{x} + \frac{1-\gamma}{2}\sigma_{l}^{y}\sigma_{l+1}^{y}\\& + \lambda\sigma_{l}^{z} + D\left( \sigma_{l}^{x}\sigma_{l+1}^{y} - \sigma_{l}^{y}\sigma_{l+1}^{x}\right)
\end{split}
\right], 
\end{equation}
where $N$ is the number of spin of the chain, $\gamma$ its anisotropy parameter, $\lambda$ the magnetude of the transverse magnetic field and D is the DM interaction which has important effects on QCs \cite{Yan et al.,You et al.,Hu,Tian et al.}. In addition, we assume the central spins, consisting of R spins, transversally coupled to the chain:
\begin{equation}
	\mathcal{H}_{I} = -\sum_{i=1}^{R}g_{i}\sigma_{i}^{z}\otimes\sum_{l=1}^{N}\sigma_{l}^{z},
	\end{equation}
	where $g_{i}$ is the coupling strength between the $i^{th}$ spin of the central system and the spin chain environment. The resulting Hamiltonian  is therefore given by:
	\begin{equation}
	\mathcal{H}_{EI} = \mathcal{H}_{E}^{\lambda} + \mathcal{H}_{I};
	\end{equation}
with $\mathcal{H}_{E}^{\lambda}$ the extension of $H_{E}^{\lambda}$ in the total Hilbert space. The canonical basis in the R-dimensional Hilbert space $ \{ |i\rangle, 0\le i \le R-1 \} $ is the eigen basis of the operator $\sum_{i=1}^{R}g_{i}\sigma_{i}^{z}$. The eigenvalue of $|i\rangle = |i_{1}i_{2}...i_{R}\rangle$ ($i_{j}\in \{0,1\}, \ \forall j \in [1,R]$) is $\alpha_{i} = \sum_{j=1}^{R}(-1)^{i_{j}}g_{j}$. So from the spectral theorem $\mathcal{H}_{EI}$ can be written as:
\begin{equation}
\label{four}
\mathcal{H}_{EI} = \sum_{k=0}^{2^{R}-1}|k\rangle\langle k|\otimes \textit{H}_{E}^{\lambda_{k}},
\end{equation}
where $ H_{E}^{\lambda_{k}} $  is obtained from Eq.~\eqref{one} by substituting $ \lambda \mbox{ by }\lambda_{k} = \lambda + \alpha_{k}. \ H_{E}^{\lambda_{k}} $ can be diagonalize following the standard procedure. We introduce the Jordan-Wigner transformation:
\begin{equation}
\left\lbrace 
\begin{split}
&\sigma_{l}^{x}=\prod_{s<l}(1-2c_{s}^{\dagger}c_{s})(c_{s}^{\dagger}+c_{s}) \\&
\sigma_{l}^{y}=-i\prod_{s<l}(1-2c_{s}^{\dagger}c_{s})(c_{s}^{\dagger}-c_{s}) \\&
\sigma_{l}^{z}=(1-2c_{l}^{\dagger}c_{l})
\end{split},
\right.
\end{equation}
to map from spin model to one-dimensional spinless fermionic model :
	\begin{equation}
\begin{split}
\textit{H}_{E}^{\lambda_{k}}=&-\sum_{l=1}^{N}\Big[(1+2iD)c_{l+1}^{\dagger}c_{l} + (1-2iD)c_{l}^{\dagger}c_{l+1}\\& + \gamma(c_{l+1}c_{l} + c_{l}^{\dagger}c_{l+1}^{\dagger}) +\lambda_{k}(1-2c_{l}^{\dagger}c_{l}) \Big],
\end{split}
\end{equation}
	where  $c_{l}^{\dagger}$ and $c_{l}$ are the creation and annihilation  operators respectively. Now, we introduce the inverse discrete Fourier transformation $c_{l} = \frac{1}{\sqrt{N}}\sum_{j=-L}^{L}d_{l}e^{\frac{2i\pi j}{N}}$ to map from real space to momentum space:
\begin{equation}
\begin{split}
\textit{H}_{E}^{\lambda_{k}}=&-\sum_{j=-L}^{L}\Bigg[
- i\gamma\sin\frac{2\pi j}{N}(d_{-j}d_{j} + d_{-j}^{\dagger}d_{j}^{\dagger}) \\& + \left\lbrace 2(\cos\frac{2\pi j}{N}-\lambda_{k}) + 4D\sin\frac{2\pi j}{N}\right\rbrace d_{j}^{\dagger}d_{j}\Bigg].
\end{split}
\end{equation}
	The diagonalized form is obtained by introducing the Bogoliubov transformation:
	\begin{equation}
	b_{j,\lambda_{k}} = \cos\left( \frac{\theta_{j}^{\lambda_{k}}}{2}\right)d_{j} - i\sin\left( \frac{\theta_{j}^{\lambda_{k}}}{2}\right) d_{-j}^{\dagger},
	\end{equation}
	with
	\begin{equation}
	\label{nine}
	\theta_{j}^{\lambda_{k}} = \arctan\left( \frac{\gamma\sin\frac{2\pi j}{N}}{\lambda_{k}-\cos\frac{2\pi j}{N}}\right), 
	\end{equation}
	as given by:
	\begin{equation}
	\textit{H}_{E}^{\lambda_{k}} = \sum_{j=-L}^{L}\Lambda_{j}^{\lambda_{k}}\left( b_{j,\lambda_{k}}^{\dagger}b_{j,\lambda_{k}} - \frac{1}{2}\right), 
	\end{equation}
	where
	\begin{equation}
	\label{eleven}
	\Lambda_{j}^{\lambda_{k}} = 2\left( \varepsilon_{j}^{\lambda_{k}} + 2D\sin\frac{2\pi j}{N}\right), 
	\end{equation}\
	and $ \varepsilon_{j}^{\lambda_{k}} $ is defined as:
	\begin{equation}
	\label{twelve}
	\varepsilon_{j}^{\lambda_{k}} = \sqrt{\left( \lambda_{k} - \cos\frac{2\pi j}{N}\right) ^2 + \gamma^2\sin^2\frac{2\pi j}{N}}.
	\end{equation}
	In addition, the normal modes $ b_{j,\lambda_{k}} $ and $ b_{j,\lambda} $ are related by:
	\begin{equation}
	 b_{j,\lambda_{k}} = \left( \cos\vartheta_{j}^{\lambda_{k}}\right) b_{j,\lambda} -i\left( \sin\vartheta_{j}^{\lambda_{k}}\right) b_{-j,\lambda}^{\dagger},
	\end{equation}
	where
	\begin{equation}
	\label{fourteen}
	\vartheta_{j}^{\lambda_{k}} = \frac{\theta_{j}^{\lambda_{k}} - \theta_{j}^{\lambda}}{2}.
	\end{equation}
	Let us assume that at $t=0$ the central spins are completely disentangled from the spin chain:
	\begin{equation}
	\rho(0) = \rho_{S}(0)\otimes|\psi_{E}\rangle\langle\psi_{E}|.
	\end{equation}
	The time evolution is given by:
	\begin{equation} \rho(t) = U(t)\rho(0)U^{\dagger}(t),
	 \end{equation}
	where
	\begin{equation}
	U(t) = \exp(-i\mathcal{H}_{EI}t)
	\end{equation}
	By tracing the total density operator on the environmental degree of freedom, the density operator of the system is obtained as:
	\begin{equation}
	\rho_{S}(t) =  \sum_{k,k'=0}^{2^{R}-1}F_{kk'}(t)\langle\phi_{k}|\rho_{S}(0)|\phi_{k'}\rangle|\phi_{k}\rangle\langle\phi_{k'}|,
	\end{equation}
	where
	\begin{equation}
	\label{nineteen}
	F_{kk'}(t) = \langle\psi_{E}|U_{E}^{\lambda_{\dagger k'}}(t)U_{E}^{\lambda_{k}}(t)|\psi_{E}\rangle,
	\end{equation}
	$ F_{kk'}(t) $ is so called decoherence factor\cite{Sun et al.,Yuan et al.,Cheng et al.}. Let us now derived the explicit expression of decoherence factor following the initial prepared state of the spin chain.
	
	\begin{widetext}
	The ground state can be chosen as:
	\begin{equation}
	\label{twenty}
	|G\rangle_{\lambda} = \prod_{j=1}^{L} \left( \cos\frac{\theta_{j}^{\lambda}}{2}|0\rangle_{j}|0\rangle_{-j} + i\sin\frac{\theta_{j}^{\lambda}}{2}|1\rangle_{j}|1\rangle_{-j}\right), 
	\end{equation}
	where, $ |1\rangle_{j}\mbox{ and }|0\rangle_{j} $ represent respectively the single excitation and the vacuum of the $j^{th}$ mode $ d_{j} $. When $ |\psi_{E}\rangle = |G\rangle_{\lambda}$, the decoherence factor reads from Eqs.~\eqref{nineteen} and \eqref{twenty} as \cite{Yuan et al.}:
	\begin{equation}
	|F_{kk'}(t)| = \prod_{j=1}^{L}\left(\Bigg| 
	\begin{split}
&\sin(\vartheta_{j}^{\lambda_{k}})\sin(\vartheta_{j}^{\lambda_{k'}})\cos(\vartheta_{j}^{\lambda_{k}}-\vartheta_{j}^{\lambda_{k'}})\times e^{-i(\Lambda_{j}^{\lambda_{k}}-\Lambda_{j}^{\lambda_{k'}})t} - \cos(\vartheta_{j}^{\lambda_{k}})\sin(\vartheta_{j}^{\lambda_{k'}}) \sin(\vartheta_{j}^{\lambda_{k}}-\vartheta_{j}^{\lambda_{k'}})e^{i(\Lambda_{j}^{\lambda_{k}}+\Lambda_{j}^{\lambda_{k'}})t}
	\\& + \sin(\vartheta_{j}^{\lambda_{k}})\cos(\vartheta_{j}^{\lambda_{k'}})\sin(\vartheta_{j}^{\lambda_{k}}-\vartheta_{j}^{\lambda_{k'}})e^{-i(\Lambda_{j}^{\lambda_{k}}+\Lambda_{j}^{\lambda_{k'}})t} + \cos(\vartheta_{j}^{\lambda_{k}})\cos(\vartheta_{j}^{\lambda_{k'}})\cos(\vartheta_{j}^{\lambda_{k}}-\vartheta_{j}^{\lambda_{k'}})e^{i(\Lambda_{j}^{\lambda_{k}}-\Lambda_{j}^{\lambda_{k'}})t}
	\end{split}
\Bigg|\right), 
	\end{equation}
	or
	\begin{equation}
|F_{kk'}(t)|=\prod_{j=1}^{L}\left(
\begin{split} 
&1-4\sin(2\vartheta_{j}^{\lambda_{k}})\sin(2\vartheta_{j}^{\lambda_{k'}})\sin^{2}(\vartheta_{j}^{\lambda_{k}}-\vartheta_{j}^{\lambda_{k'}})\sin^{2}(\Lambda_{j}^{\lambda_{k}}t)\sin^{2}(\Lambda_{j}^{\lambda_{k'}}t)
	+2\sin(2\vartheta_{j}^{\lambda_{k}})\sin(2\vartheta_{j}^{\lambda_{k'}})\sin(\Lambda_{j}^{\lambda_{k}}t)\\&\times\sin(\Lambda_{j}^{\lambda_{k'}}t)\cos(\Lambda_{j}^{\lambda_{k}}t-\Lambda_{j}^{\lambda_{k'}}t)
		-\sin^{2}(2\vartheta_{j}^{\lambda_{k}})\sin^{2}(\Lambda_{j}^{\lambda_{k}}t) -  \sin^{2}(2\vartheta_{j}^{\lambda_{k'}})\sin^{2}(\Lambda_{j}^{\lambda_{k'}}t)
\end{split}\right). 
	\end{equation}
	The vacuum state is given by:
	\begin{equation}
	|0_{E}\rangle = |0\rangle_{j=0}\otimes_{j>0}|0\rangle_{j}|0\rangle_{-j}.
	\end{equation}
	When, $ |\psi_{E}\rangle = |0_{E}\rangle$, the decoherence factor reads \cite{Sun et al.,Hu}:
	\begin{equation}
	\label{twenty-four}
	|F_{kk'}(t)| = \prod_{j=1}^{L}\left( 
	\begin{split}
	&1-\sin^{2}(\Lambda_{j}^{\lambda_{k}}t)\sin^{2}(\Lambda_{j}^{\lambda_{k'}}t)\sin^{2}(\theta_{j}^{\lambda_{k}}-\theta_{j}^{\lambda_{k'}}) - \Big[\sin(\Lambda_{j}^{\lambda_{k}}t)\cos(\Lambda_{j}^{\lambda_{k'}}t)\sin(\theta_{j}^{\lambda_{k}})	\\&- \cos(\Lambda_{j}^{\lambda_{k}}t)\sin(\Lambda_{j}^{\lambda_{k'}}t)\sin(\theta_{j}^{\lambda_{k'}})\Big] ^{2}
	\end{split}
	\right)^{\dfrac{1}{2}}.
	\end{equation}
	\end{widetext}

% If in two-column mode, this environment will change to single-column
% format so that long equations can be displayed. Use
%\begin{widetext}
%\end{widetext}

\section{DECOHERENCE PROCESS}\label{s3}
	Due to the huge difficulties to interprete the decoherence factors, we will derive a general heuristic analysis in order to understand the behaviour of decoherence process at the quantum critical point as well as far from the quantum critical point, in the weak and strong coupling and for two initial prepared states of the environment.
	\subsection{Evolution from ground state}
	\subsubsection{Weak coupling case $\left( g_{i}\ll 1, \quad \forall i \in [1,R]\right) $} Let $K_{c}$ be a cut off number, relatively small than $ N $, and $ F_{kk'} $ be its corresponding partial product of the decoherence factor:
	\begin{equation}
	|F_{kk'}(t)|_{K_{c}} = \prod_{j>0}^{K_{c}}F_{j}\geq |F_{kk'}(t)|,
	\end{equation}
	then the associated partial sum is:
	\begin{equation}
	\begin{split}
	S(t)= \ln|F_{kk'}(t)|_{K_{c}}=\ln\prod_{j>0}^{K_{c}}F_{j}=\sum_{j>0}^{K_{c}}\ln F_{j}.
	\end{split}
	\end{equation}
	When $ N $ is large enough compare to $ j $ one has from Eqs.~\eqref{eleven} and \eqref{twelve}:
	\begin{equation}
\left\lbrace 
\begin{split}
&\varepsilon_{j}^{\lambda_{k}}\approx |1-\lambda_{k}| \\&
	\Lambda_{j}^{\lambda_{k}} \approx 2|1-\lambda_{k}| + 4D\frac{2\pi j}{N}
\end{split}.
	\right.
	\end{equation}
	In addition, from Eqs.~\eqref{nine} and \eqref{fourteen}
	\begin{equation}
\left\lbrace 
	\begin{split}
	&\sin\theta_{j}^{\lambda_{k}}\approx \frac{2 \pi j\gamma}{N|\lambda_{k}-1|}\\&
		\cos\theta_{j}^{\lambda_{k}}\approx \frac{\lambda_{k}-1}{|\lambda_{k}-1|}\\&
		\sin(\vartheta_{j}^{\lambda_{k}} - \vartheta_{j}^{\lambda_{k'}})\approx-\frac{2\pi j \gamma (\lambda_{k'}-\lambda_{k})}{N|(\lambda_{k}-1)(\lambda_{k'}-1)|}
	\end{split}.
	\right.
	\end{equation}
	in the third order of $ \frac{j}{N} $, the partial sum becomes: 
	\begin{equation}
	\begin{split}
	&S(t) \approx 
		\\&\frac{1}{2}\sum_{j>0}^{K_{c}}\ln\left( 
		\begin{split}
		&1 - \frac{4\pi^{2}j^{2}\gamma^{2}\alpha_{k}^{2}\sin^{2}(\Lambda_{j}^{\lambda_{k}}t)}{N^{2}}\\& - \frac{8\pi^{2}j^{2}\gamma^{2}\alpha_{k}\alpha_{k'}\sin(\Lambda_{j}^{\lambda_{k}}t)\sin(\Lambda_{j}^{\lambda_{k'}}t) }{N^{2}(\lambda-1)^{2}|(\lambda_{k}-1)(\lambda_{k'}-1)|}\\&
		\times\cos((\Lambda_{j}^{\lambda_{k}}-\Lambda_{j}^{\lambda_{k'}})t)|(\lambda-1)(\lambda_{k}-1)|^{2} \\&- \frac{36\pi^{2}j^{2}\gamma^{2}\alpha_{k'}^{2}\sin^{2}(\Lambda_{j}^{\lambda_{k'}}t)}{N^{2}}|(\lambda_{k'}-1)(\lambda-1)|^{2}
		\end{split}
		  \right), 
	\end{split}
	\end{equation}
	for short duration we have:
	\begin{equation}
	S(t) \approx -\frac{2\pi^{2}\gamma^{2}t^{2}}{(\lambda-1)^{2}}\sum_{j>0}^{K_{c}}\left( \frac{2jt}{N}\right)^{2} \left[ \frac{\alpha_{k}\Lambda_{j}^{\lambda_{k}}}{|\lambda_{k}-1|}-\frac{\alpha_{k'}\Lambda_{j}^{\lambda_{k'}}}{|\lambda_{k'}-1|}\right] ^{2}.
	\end{equation}
	
	\subsubsection*{a. Near the quantum critical point($\lambda\rightarrow 1$)}
\begin{figure}[!htb]
	\includegraphics[scale=0.27]{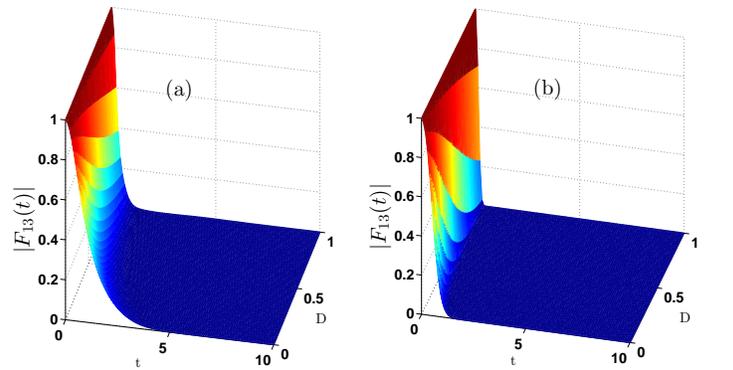}
	\caption{decoherence factor against time and DM interaction. (a) is plotted for environment prepared in the ground state and (b) in the vacuum state. N=400, g=0.05, $ \gamma=0.5 \ , \lambda=1$}
	\label{f1}
	\end{figure}

\begin{figure}[!htb]
	\includegraphics[scale=0.27]{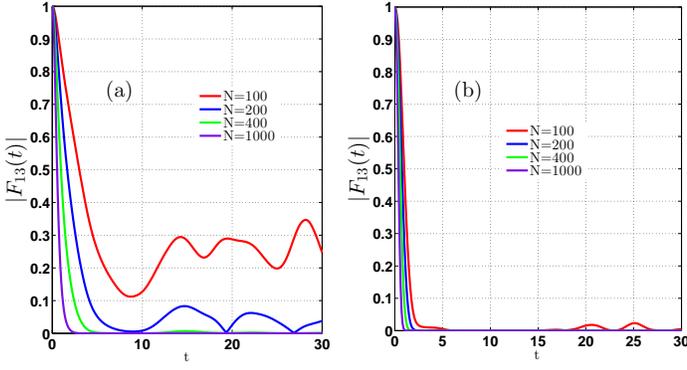}
	\caption{decoherence factor against time for many values of spin's numbers. (a) is plotted for environment prepared in the ground state and (b) for in the vacuum state. D=0, g=0.05, $ \gamma=0.5 \ , \lambda=1$}
	\label{f2}
	\end{figure}

	Near the quantum critical point, the partial sum becomes:
	\begin{equation}
	S(t) \approx -\frac{2\pi^{2}\gamma^{2}t^{2}}{(\lambda-1)^{2}}\sum_{j>0}^{K_{c}}\left( \frac{2jt}{N}\right)^{2} \left[ \frac{\alpha_{k}\Lambda_{j}^{\lambda_{k}}}{|\alpha_{k}|}-\frac{\alpha_{k'}\Lambda_{j}^{\lambda_{k'}}}{|\alpha_{k'}|}\right] ^{2}.
	\end{equation}
	In addition,$ \Lambda_{j}^{\lambda_{k}} \approx 2|\alpha_{k}| + 4D\frac{2\pi j}{N}$ thus
	\begin{equation}
	S(t) \approx - \left( \tau_{kk'}^{(1)} + \tau_{kk'}^{(2)}\right), 
	\end{equation}
	with
	\begin{equation}
\left\lbrace 
\begin{split}
	&\tau_{kk'}^{(1)} = \frac{2\gamma^{2}(\alpha_{k}-\alpha_{k'})^{2}}{(\lambda-1)^{2}}E^{(2)}(K_{c})\\&
	\tau_{kk'}^{(2)} = \frac{8D\gamma^{2}(\alpha_{k}-\alpha_{k'})}{(\lambda-1)^{2}}\left(\frac{\alpha_{k}}{|\alpha_{k}|} - \frac{\alpha_{k'}}{|\alpha_{k'}|}\right)E^{(3)}(K_{c})
\end{split},
	\right.
	\end{equation}
	and
	\begin{equation}
	E^{(i)}(K_{c}) = \sum_{j>0}^{K_{c}}\left( \frac{2\pi j}{N}\right) ^{i}.
	\end{equation}
	By maximising the decorehence factor by its partial product it becomes:
	\begin{equation}
	|F_{kk'}(t)| \approx e^{-\left( \tau_{kk'}^{(1)}  + \tau_{kk'}^{(2)}\right) t^{2}}.
	\end{equation}
	This shows that near the quantum critical point, the decoherence process will appear exponentially with the second power of time. In addition, $ E^{(2)}(K_{c})\gg E^{(3)}(K_{c})$ so $ F_{kk'}(t) $ is mainly determined by $ \tau_{kk'}^{(1)} $ and then the DM interaction has a slightly effect on the decoherence process in particular. These are consistent with the numerical result of exact expression of decoherence factor in Fig.~\eqref{f1}(a). As $ E^{(2)}(K_{c}) $ and $ E^{(3)}(K_{c}) $ are inversely proportional to N and consequently to $\tau_{kk'}^{(1)}$ and $\tau_{kk'}^{(2)}$, then, this implies that, the higher the number of spins in the chain the faster the decoherence process. This result is consistent with the numerical result of Fig.~\eqref{f2}(a). The results of \cite{Yuan et al.,Cheng et al.} can be derived from those obtained here by setting $k=1$ and $k'=3$. On the other hand, when $ \gamma \rightarrow 0 \mbox{ then } \tau_{1}, \tau_{2} \rightarrow 0$ and as a result $ |F_{kk'}(t)| \rightarrow 1$ therefore, the XX model will not affect the central spins system when the system is weakly coupled to the environment at the quantum critical point.

	\subsubsection*{b. Far the quantum critical point($\lambda\gg1$)} 
\begin{figure}[!htb]
		\includegraphics[scale=0.27]{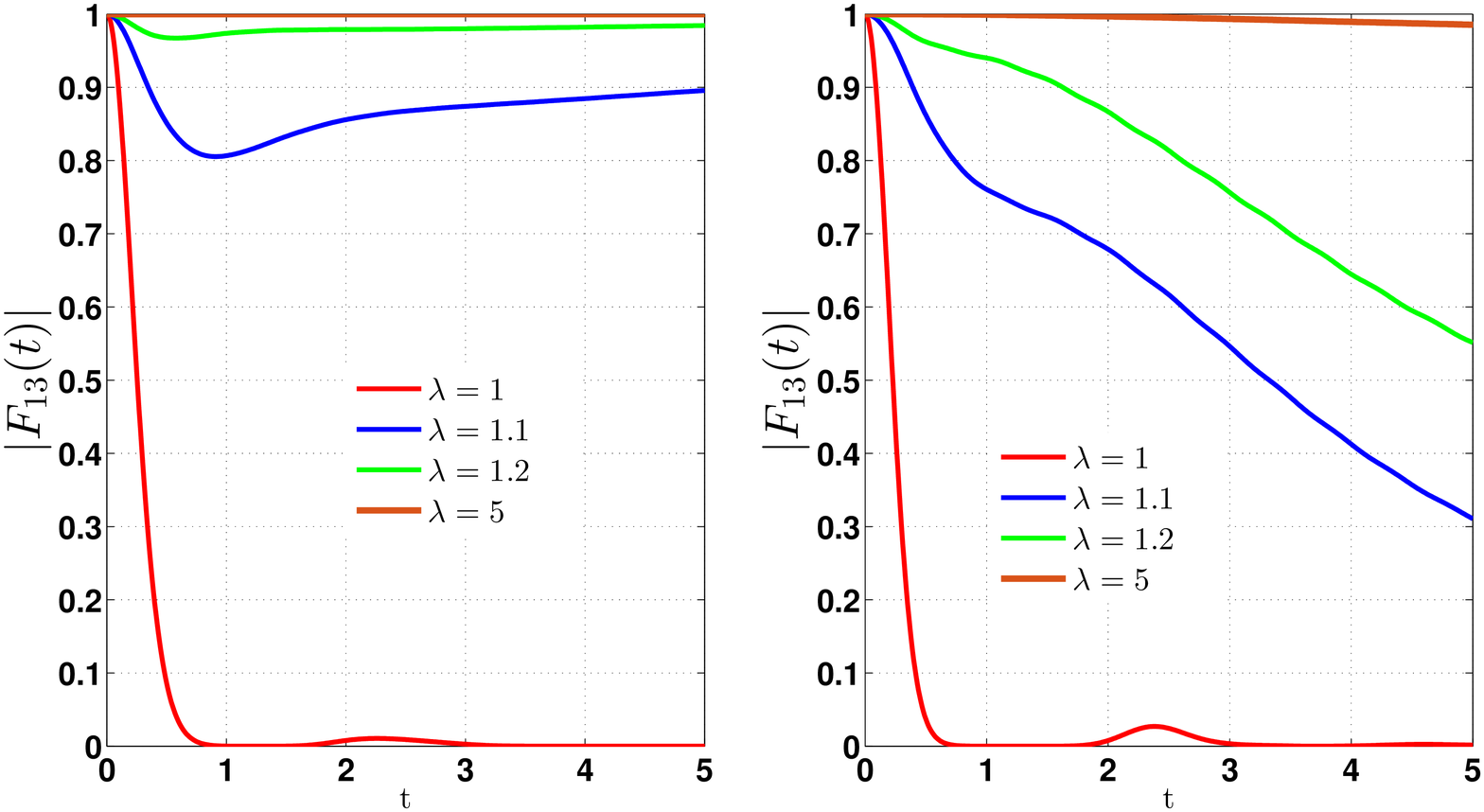}
		\caption{decoherence factor against time for different magnetics fields . (a) is plotted for environment prepared in the ground state and (b) for in the vacuum state. D=1, N =400 , g=0.05, $ \gamma=0.1 $}
		\label{f3}
	\end{figure}

When we are relatively far from the quantum critical point such that
	\begin{equation}
	|\lambda_{k} - 1| \approx |\lambda - 1|,\,\forall k
	\end{equation}
	namely, as $ g_{i} \rightarrow 0 $ then
	\begin{equation}
	g_{i} \ll |\lambda - 1|
	\end{equation}			
	Thus the partial sum at the third power of $ \frac{j}{N} $ becomes
	\begin{equation}
	S(t) \approx - \left( \tau_{kk'}^{(1)} + \tau_{kk'}^{(2)'}\right), 
	\end{equation}
	with
	\begin{equation}
	\tau_{kk'}^{(2)'} = \frac{8D\gamma^{2}(\alpha_{k'}-\alpha_{k})^{2}}{(\lambda-1)^{2}}E^{(3)}(K_{c}).
	\end{equation}
	By maximising the decorehence factor by its partial product it comes that
	\begin{equation}
	|F_{kk'}(t)| \approx e^{-\left( \tau_{kk'}^{(1)}  + \tau_{kk'}^{(2)'}\right)t^{2}}.
	\end{equation}
	By comparing $\tau_{kk'}^{(2)}$ and $\tau_{kk'}^{(2)'}$ it comes that: far from the quantum critical point the decoherence process is weaker than at the quantum critical point, this is consistent with the numerical result of ref.\cite{Yuan et al.}. We relate this to the fact that strong magnetic field disturbed the spin chain, such that far from the quantum critical point, the coherence of the spins of the environment becomes weak, contrary at the quantum critical point where the coherence of the spins of the environment is maximal and they collaborate very well. The result is the vanishing coherence of the central spins system; which is consistent with the numerical result of Fig.~\eqref{f3}(a) where the exact expression of decoherence factor is plotted. We have also the same results for D and N as in the case of quantum critical point. In addition, as $\tau_{kk'}^{(2)} \gg \tau_{kk'}^{(2)'}$, the farther the quantum critical point, the weaker is the effect of DM interaction on decoherence process.
	
%	\begin{figure}[!htb]
%		\centering
%		\includegraphics[scale=0.5]{Figures/WeakFarTDm.png}
%		\caption{decoherence factor against time and DM interaction. (a) is plotted for environment prepared in the ground state and (b) for environment  prepared in the vacuum state. N=400, g=0.05, $ \gamma=0.5 \ , \lambda=1.1$ \label{WeakFarTDm}}
%	\end{figure}
	
	\subsubsection{Strong coupling case.} When $g_{i}\gg 1 \ \forall i$ then $ \lambda_{k} \rightarrow \alpha_{k}(+\infty) $ thus,
	\begin{equation}
\left\lbrace 
\begin{split}
	&\tan\theta_{j}^{\lambda_{k}}\rightarrow 0^{+}\text{if}\quad \alpha_{k} > 0 \\&
	\tan\theta_{j}^{\lambda_{k}}\rightarrow 0^{-}\text{ if}\quad \alpha_{k} < 0
\end{split},
	\right.
	\end{equation}
	therefore,
	\begin{equation}
\left\lbrace 
\begin{split}
	&\theta_{j}^{\lambda_{k}}\rightarrow 0 \text{ if } \alpha_{k} > 0 \\&
	\theta_{j}^{\lambda_{k}}\rightarrow \pi \text{ if } \alpha_{k} < 0
\end{split}.
	\right.
	\end{equation}
	\subsubsection*{a. When $ \alpha_{k}>0 $ and $ \alpha_{k'}<0 $ or $ \alpha_{k}<0 $ and $ \alpha_{k'}>0 $} 
\begin{figure}[!htb]
		\includegraphics[scale=0.27]{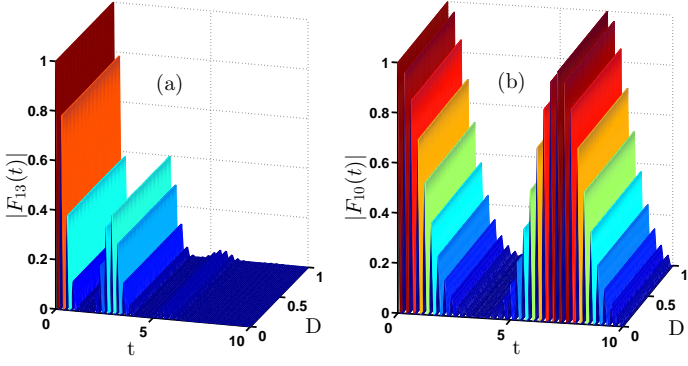}
		\caption{decoherence factor against time and DM interaction in ground state. (a) is plotted for $\alpha_{k}\alpha_{k'}<0$ and (b) for $ \alpha_{k}\alpha_{k'}>0 $. N=400, g=500, $ \gamma=0.5 \ , \lambda=1.1$}
		\label{f4}
	\end{figure}

As the result for the two sets are the same and from the symmetry of the decoherence factor, we will derive the heuristic analysis from the first set.
	Here we have
	\begin{equation}
\left\lbrace 
\begin{split}
&\theta_{j}^{\lambda_{k}} \approx 0 \ ,\quad \theta_{j}^{\lambda_{k'}} \approx \pi \\&
\vartheta_{j}^{\lambda_{k}} - \vartheta_{j}^{\lambda_{k'}} \approx -\frac{\pi}{2}
\end{split},
\right.
	\end{equation}
	with these approximations, the decoherence factor becomes as follow \cite{Cucchietti,Cheng et al.},
	\begin{equation}
	\begin{split}
	|F_{kk'}(t)|&\approx\prod_{j>0}|\cos^{2}\frac{\theta_{j}^{\lambda}}{2}e^{i(\Lambda_{j}^{kk'})t} + \sin^{2}\frac{\theta_{j}^{\lambda}}{2}e^{-i(\Lambda_{j}^{kk'})t}| \\&
		=\prod_{j>0}|\cos((\Lambda_{jkk'})t) + i\sin((\Lambda_{jkk'})t)\cos\frac{\theta_{j}^{\lambda}}{2}| \\&
		\approx e^{\frac{-(S_{kk'}^{2}t^{2})}{2}}|\cos((\Lambda_{kk'})t)|^{\frac{N}{2}}
	\end{split},
	\end{equation}
	where $ \Lambda_{kk'} $ is given by:
	\begin{equation}
	\begin{split}
	\Lambda_{kk'}&=\frac{1}{L}\sum_{j>0}(\Lambda^{\lambda_{k}}_{j} + \Lambda^{\lambda_{k'}}_{j}) \\&
		\approx\frac{2}{L}\sum\Big[\left( \alpha^{2}_{k} + \gamma^{2}\sin^{2}{\frac{2\pi j}{N}}\right)^{\frac{1}{2}} \\&+  \left( \alpha^{2}_{k'} + \gamma^{2}\sin^{2}{\frac{2\pi j}{N}}\right)  ^{\frac{1}{2}} + 4D\sin\left( \frac{2\pi j}{N}\right)\Big]\\&
		\approx 2(|\alpha_{k}| + |\alpha_{k'}|) + \frac{\gamma^{2}}{2}\left( \frac{1}{|\alpha_{k}|} + \frac{1}{|\alpha_{k'}|}\right) 
	\end{split}.
	\end{equation}
	In addition,
	\begin{equation}
	S_{N}^{2} = \sum_{j>0}\sin^{2}(\theta_{j}^{\lambda})\delta_{j}^{2},
	\end{equation}
	with $ \delta_{j} $ representing the deviation of $ \Lambda_{jkk'} $ from its mean values:
	\begin{equation}
	\delta_{j} = \Lambda_{jkk'} - \Lambda_{kk'}, 
	\end{equation}
	and
	\begin{equation}
	\begin{split}
	&\Lambda_{kk'} = \Lambda_{j}^{k} + \Lambda_{j}^{k'}\\&
		\approx 2\left( |\alpha_{k}| + |\alpha_{k'}|\right)+\gamma^{2}\left( \frac{1}{|\alpha_{k}|} + \frac{1}{|\alpha_{k'}|}\right) \sin^{2}\left( \frac{2/pi j}{N}\right)  \\&+ 8D\sin\left( \frac{2\pi j}{N}\right) 
	\end{split},
	\end{equation}
	then
	\begin{equation}
	\delta_{j}\approx 8D\sin\left( \frac{2\pi j}{N}\right)  - \frac{\gamma^{2}}{2}\left( \frac{1}{|\alpha_{k}|} + \frac{1}{|\alpha_{k'}|}\right) \cos\left( \frac{4\pi j}{N}\right). 
	\end{equation}
	In this configuration, the decoherence process exhibits an oscillatory Gaussian envelope with the period 
	\begin{equation}
		P = \pi\left( 2\left( |\alpha_{k}| + |\alpha_{k'}|\right)  + \frac{\gamma^{2}}{2}\left( \frac{1}{|\alpha_{k}|} + \frac{1}{|\alpha_{k'}|}\right)\right)^{-1},
	\end{equation} 
	and width 
	\begin{equation}
		W = \left\lbrace \left( \frac{\gamma^{4}}{4}\left(\frac{1}{|\alpha_{k}|}+\frac{1}{|\alpha_{k'}|}\right) ^{2} + 64D^{2}\right) N\right\rbrace ^{-\frac{1}{2}}.
	\end{equation} 
	It is important to note that W represents also the decoherence time. In addition, strong DM interaction will enhance the decoherence process by increasing the envelope width as shown in the numerical result of exact expression in Fig.~\eqref{f4}(a). As $|F_{kk'}(t)|$ is independent of $\lambda$, the decoherence process is independent of the position from the quantum critical point. The results of \cite{Yuan et al.,Cheng et al.} can be derived from those obtained here by setting $k=1$ and $k'=3$.
	\subsubsection*{b. When $ \alpha_{k}>0 $ and $ \alpha_{k'}>0 $ or $ \alpha_{k}<0 $ and $ \alpha_{k'}<0 $}
\begin{figure}[!htb]
		\includegraphics[scale=0.27]{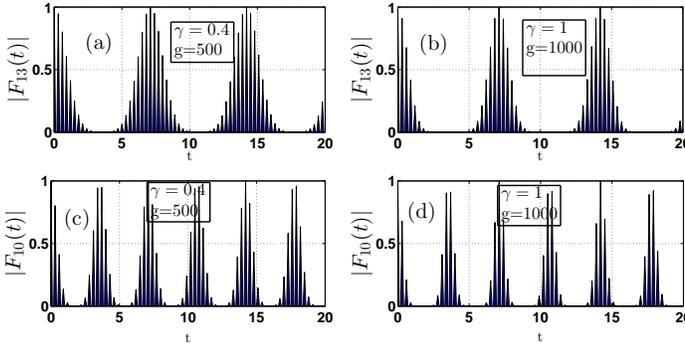}
		\caption{decoherence factor against time in ground state for $\alpha_{k}\alpha_{k'}>0$. D=0.5, N =400 , $ \lambda=1 $}
		\label{f5}
	\end{figure}

	 As the result for the two sets are the same because of the symmetry of the decoherence factor, we will derive the heuristic analysis from the first set. In this case we have:
	\begin{equation}
\left\lbrace 
\begin{split}
&\theta_{j}^{\lambda_{k}} = \theta_{j}^{\lambda_{k'}}\rightarrow 0 \\&
\vartheta_{j}^{\lambda_{k}} - \vartheta_{j}^{\lambda_{k'}}\rightarrow 0
\end{split},
\right.
	\end{equation}
	thus the same argumentation as the previous one gives
	\begin{equation}
	\begin{split}
	&\Lambda_{kk'}= \frac{1}{L}\sum_{j>0}\left( \Lambda^{\lambda_{k}}_{j} - \Lambda^{\lambda_{k'}}_{j}\right)\\&
		\approx 2\left(|\alpha_{k}| - |\alpha_{k'}|\right)  + \frac{\gamma^{2}}{2}\left( \frac{1}{|\alpha_{k}|} - \frac{1}{|\alpha_{k'}|}\right) 
	\end{split},
	\end{equation}
	and
	\begin{equation}
	\Lambda_{kk'} \approx 2\left(|\alpha_{k}| - |\alpha_{k'}|\right) + \gamma^{2}\left( \frac{1}{|\alpha_{k}|} - \frac{1}{|\alpha_{k'}|}\right) \sin^{2}\left( \frac{2/pi j}{N}\right), 
	\end{equation}
	then
	\begin{equation}
	\delta_{j} \approx 0 \quad \Rightarrow \quad S_{N}^{2} \approx 0,
	\end{equation}
	therefore
	\begin{equation}
	|F_{kk'}(t)| \approx |\cos((\Lambda_{kk'})t)|^{\frac{N}{2}}.
	\end{equation}
	In this situation, the decoherence process exhibits an oscillatory behaviour with the period
	\begin{equation}
		P =  \pi\left( 2\left( |\alpha_{k}| - |\alpha_{k'}|\right)+\frac{\gamma^{2}}{2}\left( \frac{1}{|\alpha_{k}|} - \frac{1}{|\alpha_{k'}|}\right) \right)^{-1}.
	\end{equation}
	 This is consistent with the numerical result of Fig.~\eqref{f5}. As $|F_{kk'}(t)|$ is independent to D and $\lambda$, the DM interaction has no effect on the decoherence process and the decoherence process is independent to the position from the quantum critical point. This is consistent with numerical results of Fig.~\eqref{f5}(b). In addition, as for larger N, $|F_{kk'}(t)|$ is small, we can conclude that the higher the number of spin in the chain, the faster the decoherence process.

	\subsection{Evolution from vacuum state}
	\subsubsection{Strong coupling case.} 
\begin{figure}[!htb]
		\includegraphics[scale=0.27]{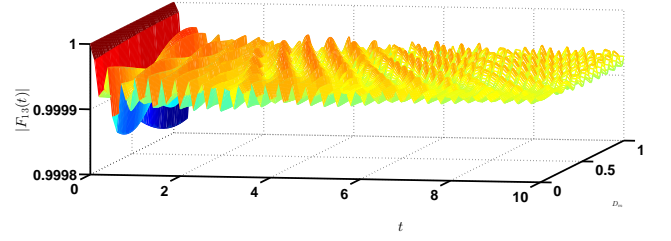}
		\caption{decoherence factor against time and DM interaction in vacuum state. N=400, g=500, $ \gamma=0.5 \ , \lambda=1$}
		\label{f6}
	\end{figure}

When $ g_{i}\gg 1  \ \forall i$ we have $ \lambda_{k}\rightarrow \alpha_{k}(+\infty) $ , thus it comes that:
	\begin{equation}
\left\lbrace 
\begin{split}
&\frac{\gamma sin\left( \frac{2\pi j}{N}\right) }{\lambda_{k} -\cos\left( \frac{2\pi j}{N}\right) } \rightarrow  0^{+} \text{ if } \alpha_{k}>0 \\&
	\frac{\gamma sin\left( \frac{2\pi j}{N}\right) }{\lambda_{k}-\cos\left( \frac{2\pi j}{N}\right) } \rightarrow  0^{-} \text{ if } \alpha_{k}<0
\end{split}.
	\right.
	\end{equation}
	Therefore,
	\begin{equation}
\left\lbrace 
\begin{split}
&\theta_{j}^{\lambda_{k}} = 0 \mbox{ if } \alpha_{k}>0\\&
	\theta_{j}^{\lambda_{k}} = \pi \mbox{ if } \alpha_{k}<0
\end{split}.\right.
	\end{equation}
	Thus, from Eq.~\eqref{twenty-four} it comes that:  
	\begin{equation}
	|F_{kk'}(t)| \approx 1.
	\end{equation}
	Therefore, when the spin chain is prepared in the vacuum state and is strongly coupled to the central spins, decoherence process is independent from the physical parameters especially on the position from the quantum critical point, and will not hold. Thus the central spins will conserve their quantum coherences. This result is consistent with the particular one of the literature \cite{Hu} and also with the numerical result of the exact expression of decoherence factor in Fig.~\eqref{f6}.
	
	\subsubsection{Weak coupling case ($g \ll 1$)} Here, the same argumentation as what have been done in the case of evolution from the ground state gives the following partial sum:
	\begin{equation}
\begin{split}
	&S(t)=-\frac{1}{2}\sum_{j>0}\ln\Bigg( 1- \left( \frac{2\pi j\gamma(\lambda_{k'}-\lambda_{k})}{N|(\lambda_{k'}-1)(\lambda_{k}-1)|}\right) ^{2}\\& \times\sin^{2}(\Lambda_{j}^{\lambda_{k}}t)\sin^{2}(\Lambda_{j}^{\lambda_{k'}}t) -\left( \frac{2\pi j\gamma}{N}\right) ^{2} \\&\times\left[ \frac{\sin(\Lambda_{j}^{\lambda_{k}}t)\cos(\Lambda_{j}^{\lambda_{k'}}t)}{|(\lambda_{k}-1)|} - \frac{\sin(\Lambda_{j}^{\lambda_{k'}}t)\cos(\Lambda_{j}^{\lambda_{k}}t)}{|(\lambda_{k'}-1)|}\right] ^{2}\Bigg) 
\end{split}.
	\end{equation}
	As for small $ x,\quad \ln(1+x)\approx x $, we have in the short time limit and at the seventh order of the product of $ t $ and $ \frac{j}{N} $
\begin{equation}
\begin{split}
&S(t) = -\frac{1}{2}\sum_{j>0}\Bigg( 16t^{4}(\alpha_{k'}-\alpha_{k})^{2}\left( \frac{2\pi j\gamma}{N}\right) ^{2} \\&
	+ 16(tD\gamma)^{2}\left( \frac{2\pi j\gamma}{N}\right)^{4}\left[ \frac{1}{|\lambda_{k}-1|} -\frac{1}{|\lambda_{k'}-1|}\right] ^{2}\\&+ 64D\gamma^{2}t^{4}(\alpha_{k'}-\alpha_{k})^{2}\left(\frac{2\pi j\gamma}{N}\right)^{3}\left[\frac{1}{|\lambda_{k}-1|} +\frac{1}{|\lambda_{k'}-1|}\right] \Bigg)
\end{split}.
\end{equation}
	\subsubsection*{a. Near the quantum critical point} In the vicinity of the quantum critical point, by maximising the decoherence factor by its partial product it comes that:
	\begin{equation}
	|F_{kk'}(t)| \approx e^{-\left(\tau^{''(1)}_{kk'} + \tau^{''(2)}_{kk'}\right)t^{4}}e^{-\left(\tau^{''(3)}_{kk'}\right)t^{2}}
	\end{equation}
	where: 
	\begin{equation}
\left\lbrace 
\begin{split}
&\tau_{kk'}^{''(1)}=8\gamma^{2}(\alpha_{k'}-\alpha_{k})^{2}E_{K_{c}}^{(2)} \\&
	\tau_{kk'}^{''(2)}=32D\gamma^{2}(\alpha_{k'}-\alpha_{k})^{2}E_{K_{c}}^{(3)}\left[ \frac{1}{|\alpha_{k}|} +\frac{1}{|\alpha_{k'}|}\right]  \\&
	\tau_{kk'}^{''(3)}=8D^{2}\gamma^{2}E_{K_{c}}^{(4)}\left[ \frac{1}{|\alpha_{k}|} -\frac{1}{|\alpha_{k'}|}\right] ^{2}
\end{split}.
\right.
	\end{equation}
	Therefore, in the case of evolution from vacuum state, at the quantum critical point, as $ E^{(2)}(K_{c})\gg E^{(3)}(K_{c})$ and $ E^{(2)}(K_{c})\gg E^{(4)}(K_{c})$ so $ F_{kk'}(t) $ is mainly determined by $ \tau_{kk'}^{''(1)} $ and the DM interaction has a slightly effect on decoherence process; what is consistent with the numerical result of Fig.~\eqref{f1}(b). In addition, as $ E^{(2)}(K_{c}) ,\ E^{(3)}(K_{c}) \mbox{ and } E^{(4)}(K_{c})$ are proportional to N therefore, the greater the number of spins in the chain the faster the decoherence process, this is also consistent with the numerical result of the exact expression in Fig.~\eqref{f2}(b). By comparing Fig.~\eqref{f2}(a) and (b), it comes that the spin chain prepared in the ground state is more sensitive to the variation of the number of spins in the chain than the one prepared in the vacuum state. When $ \gamma \rightarrow 0 , \mbox{ then } \tau_{kk'}^{''(1)}, \tau_{kk'}^{''(2)}, \tau_{kk'}^{''(3)} \rightarrow 0$ and as a result $ |F_{kk'}(t)| \rightarrow 1$, hence the XX model will not affect the central spins system when the system is weakly coupled to the environment at the quantum critical point. In addition,
	when $t\ll \frac{j}{N}$ then,
	\begin{equation}
	|F_{kk'}(t)|\approx e^{-\left(\tau^{''(3)}_{kk'}\right)t^{2}},
	\end{equation}
	hence, decoherence process appears exponentially with the second power of time and are slower that the one induced when the environment is prepared in the ground state. On the other hand, when $t\gg \frac{j}{N}$ then,
	\begin{equation}
	|F_{kk'}(t)| \approx e^{-\left(\tau^{''(1)}_{kk'} + \tau^{''(2)}_{kk'}\right)t^{4}}.
	\end{equation}
	The decoherence process will exponentially appear with the fourth power of time contrary to the case of spin chain prepare in ground state, where it appears exponentially with second power of time. The same result was obtained when $k=1$ and $k'=3$, by Sun et al. \cite{Sun et al.} and Hu \cite{Hu}.
	
	\subsubsection*{b. Far the quantum critical point ($\lambda \gg 1$)} When we are not near the quantum critical point, that is relatively far from the quantum critical point such that the following relation holds:
	\begin{equation}
	|\lambda - 1| \gg g,
	\end{equation}
	then,
	\begin{equation}
	\frac{1}{|\lambda_{k} - 1|} \approx \frac{1}{|\lambda - 1|}.
	\end{equation}
	Thus, by maximising the decoherence factor by its partial product we have
	\begin{equation}
	|F_{kk'}(t)| \approx e^{-\left( \tau^{''(1)}_{kk'} + \tau^{'''(2)}_{kk'}\right)t^{4}},
	\end{equation}
	where: 
	\begin{equation}
	\tau_{kk'}^{'''(2)} = \frac{64D\gamma^{2}(\alpha_{k'}-\alpha_{k})^{2}E_{K_{c}}^{(3)}}{|\lambda-1|}.	
	\end{equation}
	Therefore, as far from the quantum critical point $ \frac{|\lambda-1|}{g_{i}} \gg 1 $, then the farther the quantum critical point, the weaker the decoherence process. Once more, we relate this to the fact that strong magnetic field disturbed the spin chain, such that far from the quantum critical point, the coherence of the spins of the environment becomes weak, contrary at the quantum critical point where the coherence of the spins of the environment is maximal and they collaborate very well. The result is the vanishing coherence of the central spins system; which is consistent with the numerical result of Fig.~\eqref{f3}(b) where the exact expression of decoherence factor is plotted. We have also the same results about D and N as in the case at quantum critical point. In addition, as $\tau_{kk'}^{''(2)} \gg \tau_{kk'}^{'''(2)}$, the farther the quantum critical point, the stronger is the effect of DM interaction on decoherence process.
	
	\section{Applications: Quantum Correlations dynamics}\label{s4}
	We are here going to investigate the QCs dynamics measure by the tripartite Negativity and the  Genuine Tripartite Quantum Discord (GTQD) of three interacting qubits under the previous decoherence process. In particular, we set R to three and model the interaction between the three-qubit system as the XXZ Heisenberg model with DM interaction:
	\begin{equation}
	\begin{split}
	H_{S}&=\frac{J}{2}\Big[ \sigma_{1}^{x}\sigma_{2}^{x} + \sigma_{1}^{y}\sigma_{2}^{y} + \Delta\sigma_{1}^{z}\sigma_{2}^{z} + \sigma_{1}^{x} \sigma_{3}^{x} + \sigma_{1}^{y} \sigma_{3}^{y} + \Delta\sigma_{1}^{z}\sigma_{3}^{z} \\&
		+\sigma_{2}^{x}\sigma_{3}^{x} + \sigma_{2}^{y}\sigma_{3}^{y} + \Delta \sigma_{2}^{z}\sigma_{3}^{z} + M\big( \sigma_{1}^{x}\sigma_{2}^{y} - \sigma_{1}^{y}\sigma_{2}^{x} + \sigma_{1}^{y}\sigma_{3}^{x} \\&
		-\sigma_{1}^{x}\sigma_{3}^{y} + \sigma_{2}^{x}\sigma_{3}^{y}  - \sigma_{2}^{y}\sigma_{3}^{x}\big)\Big] + B\left(\sigma_{1}^{z} +\sigma_{2}^{z} + \sigma_{3}^{z}\right) 
	\end{split},
	\end{equation}
	where $ J $ is the coupling constant between spins of central system, $\Delta$ the anisotropy parameter in the z-direction and M the z-component of the DM interaction. The Hamiltonian of the total system can be written as:
	\begin{equation}
	\mathcal{H} = \mathcal{H}_{S} + \mathcal{H}_{EI},
	\end{equation}
	where $\mathcal{H}_{S}$ is the extension of $H_{S}$ in the total Hilbert space. The Diagonalization of $H_{S}$ gives the following eigen energies:
	\begin{equation}
	\label{seventy-three}
\left\lbrace 
\begin{split}
&E_{0}= 3\lambda+\frac{3J\Delta}{2} \\&
	E_{1} = -J - JM\sqrt{3} + \lambda - \frac{J\Delta}{2} \\&
	E_{2} = -J + JM\sqrt{3} + \lambda - \frac{J\Delta}{2} \\&
	E_{3} = 2J - \lambda - \frac{J\Delta}{2} \\&
	E_{4} = 2J + \lambda - \frac{J\Delta}{2} \\&
	E_{5} = -J + JM\sqrt{3} - \lambda - \frac{J\Delta}{2} \\&
	E_{6} = -J - JM\sqrt{3} - \lambda - \frac{J\Delta}{2} \\&
	E_{7} = -3\lambda+\frac{3J\Delta}{2}
\end{split}.
	\right.
	\end{equation}
	From which the corresponding eigen states are:
	\begin{equation}
\left\lbrace 
\begin{split}
&|\phi_{0}\rangle = |000\rangle \\&
	|\phi_{1}\rangle = \frac{1}{\sqrt{3}}(e^{i\frac{\pi}{6}}|000\rangle + e^{i\frac{5\pi}{6}}|010\rangle + e^{i\frac{9\pi}{6}}|100\rangle) \\&
	|\phi_{2}\rangle = \frac{1}{\sqrt{3}}(e^{i\frac{5\pi}{6}}|001\rangle + e^{i\frac{\pi}{6}}|010\rangle + e^{i\frac{9\pi}{6}}|100\rangle) \\&
	|\phi_{3}\rangle = \frac{1}{\sqrt{3}}(|011\rangle + |101\rangle + |110\rangle) \\&
	|\phi_{4}\rangle = \frac{1}{\sqrt{3}}(|001\rangle + |010\rangle + |100\rangle) \\&
	|\phi_{5}\rangle = \frac{1}{\sqrt{3}}(e^{i\frac{5\pi}{6}}|011\rangle + e^{i\frac{\pi}{6}}|101\rangle + e^{i\frac{9\pi}{6}}|110\rangle) \\&
	|\phi_{6}\rangle = \frac{1}{\sqrt{3}}(e^{i\frac{\pi}{6}}|011\rangle + e^{i\frac{5\pi}{6}}|101\rangle + e^{i\frac{9\pi}{6}}|110\rangle) \\&
	|\phi_{7}\rangle =|111\rangle
\end{split}.
	\right.
	\end{equation}
	We obtain the commutation relations
	\begin{equation}
	\label{seventy-five}
	[\textit{H}_{S}, g(\sigma_{1}^{z} + \sigma_{2}^{z} + \sigma_{3}^{z})] = 0,
	\end{equation}
	and
	\begin{equation}
	[\mathcal{H}_{S},\mathcal{H}_{EI}] = 0.
	\end{equation}
	From Eqs.~\eqref{four}, \eqref{seventy-three} and \eqref{seventy-five} $\mathcal{H}_{EI}$ reads as follow:
	\begin{equation}
	\mathcal{H}_{EI} = \sum_{k=0}^{7}|\phi_{k}\rangle\langle\phi_{k}|\otimes \textit{H}_{E}^{\lambda_{k}}.
	\end{equation}
	So the density operator of the central spin is now given by:
	\begin{equation}
	\rho_{S}(t) =  \sum_{k,k'=0}^{7}e^{-i(E_{k}-E_{k'})t}F_{kk'}(t)\langle\phi_{k}|\rho_{S}(0)|\phi_{k'}\rangle|\phi_{k}\rangle\langle\phi_{k'}|.
	\end{equation}
	Let us assume the three-qubit system initially prepared in a mixed state composed of a W and GHZ state, namely,
	\begin{equation}
	|\psi_{S}(0)\rangle = a|GHZ\rangle + \sqrt{1-a^{2}}|W\rangle,
	\end{equation}
	where, $a \ \in \{0, 1\}$, $|GHZ\rangle = \frac{1}{\sqrt{2}}(|000\rangle + |111\rangle) ,\ |W\rangle = \frac{1}{\sqrt{3}}(|001\rangle + |010\rangle + |100\rangle)$. So the time dependent density matrix can be written as follow:
	\begin{equation}
	\rho_{S}(t) = \begin{pmatrix}
	\frac{a^{2}}{2} & 0 & 0 & 0 & 0 & 0 & 0 & \epsilon \\
	0 & 1-a^{2} & 1-a^{2} & 0 & 1-a^{2} & 0 & 0 & 0 \\
	0 & 1-a^{2} & 1-a^{2} & 0 & 1-a^{2} & 0 & 0 & 0 \\
	0 & 0 & 0 & 0 & 0 & 0 & 0 & 0 \\
	0 & 1-a^{2} & 1-a^{2} & 0 & 1-a^{2} & 0 & 0 & 0 \\
	0 & 0 & 0 & 0 & 0 & 0 & 0 & 0 \\
	0 & 0 & 0 & 0 & 0 & 0 & 0 & 0 \\
	\epsilon^{*} & 0 & 0 & 0 & 0 & 0 & 0 & \frac{a^{2}}{2}
	\end{pmatrix},
	\end{equation}
	where $\epsilon = \frac{a^{2}}{2}F_{07}e^{-i(E_{0}-E_{4})t}$
	
	We use the \emph{tripartite Negativity} as the measure of Entanglement of the three-qubit system. The tripartite Negativity is define as \cite{Kenfack et al.}:
	\begin{equation}
	\mathcal{N}^{(3)}(\rho_{ABC}(t)) = \sqrt[3]{\mathcal{N}_{A-BC}\mathcal{N}_{B-AC}\mathcal{N}_{C-AB}}, 
	\end{equation}
	which represents the geometric mean of the bipartite negativity $ \mathcal{N}_{I-JK} $. As our state is symmetric, all the bipartite negativity are equal. Thus the tripartite negatibity $ \mathcal{N}^{(3)} $ reduces to the bipartite negativity of any bipartition of our system, namely \cite{Kenfack et al.}:
	\begin{equation}
\begin{split}
\mathcal{N}^{(3)}(\rho_{ABC}(t))&= \mathcal{N}_{C-AB} \\&
	= \sum_{i=0}^{7}|a_{i}(\rho_{ABC}^{T_{C}}(t))| - 1
\end{split},
	\end{equation}
where $ \rho_{ABC}^{T_{C}}(t) $ is the \emph{partial transpose} of the density matrix with respect to the subsystem C, and $ a_{i}(\rho_{ABC}^{T_{C}}(t)) $ are the eigenvalues of $ \rho_{ABC}^{T_{C}}(t) $.
After calculation, the tripartite Negativity reads as:
	\begin{equation}
	\begin{split}
	\mathcal{N}^{(3)}(\rho_{ABC}(t))&= -\frac{a^{2}+2}{6} + \sqrt{\frac{a^{4}}{4} + 8\left( \frac{1-a^{2}}{3}\right) ^{2}} \\&+\sqrt{
			a^{4}|F_{07}|^{2} + \left( \frac{1-a^{2}}{3}\right) ^{2}}
	\end{split}.
	\end{equation}
	The GTQD, $ \mathcal{D}^{(3)}(\rho) $, is used as a measure of Quantum Discord. It is given by \cite{Kenfack et al.}:
	\begin{equation}
	\mathcal{D}^{(3)}(\rho_{ABC}) = \mathcal{T}^{3}(\rho_{ABC}) - \mathcal{J}^{3}(\rho_{ABC}),
	\end{equation}
	where $ \mathcal{T}^{3}(\rho) $ is the genuine total correlations and $ \mathcal{J}^{3}(\rho) $ is the genuine classical correlation. As the state is symmetric we have:
	\begin{equation}
\left\lbrace 
	\begin{split}
	&\mathcal{T}^{3}(\rho_{ABC}) = S(\rho_{C}) + S(\rho_{AB}) - S(\rho_{C}) \\&
	\mathcal{J}^{3}(\rho_{ABC}) = S(\rho_{C}) - S(\rho_{C|AB})
	\end{split},
	\right.
	\end{equation}
	where $ S(\rho) = -Tr(\rho\log(\rho))$ is the von Neumann entropy and $S(\rho_{C|AB}) = \min_{E_{ij}^{AB}}\sum_{i,j=1}^{2}p_{ij}S(\rho_{C|E_{ij}^{AB}})$ is the quantum conditional entropy. After a tedious calculation following the method described in ref.\cite{Beggi et al}, the GTQD reads as follows:
\begin{equation}
\begin{split}
	&\mathcal{D}^{(3)}(\rho_{ABC}(t))=  \left( \frac{a}{2}\left( 1-|F_{07}|\right) \right) \log\left( 1-|F_{07}|\right)  \\&+\left( \frac{a}{2}\left( 1+|F_{07}|\right) \right) \log\left( 1+|F_{07}|\right)  - \frac{a}{2}\\&-\frac{2+a^{2}}{6}\log\left( \frac{2+a^{2}}{6}\right)-\frac{2(1-a^{2})}{3}\log\left( \frac{2}{3}\right) 
\end{split}.
\end{equation}
	When the system is prepared in the W state, namely, when $a=0$ the QCs read as:
	\begin{equation}
\left\lbrace 
\begin{split}
&\mathcal{N}^{(3)}(\rho_{ABC}(t)) = \frac{2\sqrt{2}}{3}\\&
\mathcal{D}^{(3)}(\rho_{ABC}(t)) = \log 3 - \frac{2}{3}
\end{split}.
	\right.
	\end{equation}
	As the QCs are conserved here, the coherence between pointer states are conserved, this implies that the W state is already a pointer state of the spin chain as apparatus following the quantum measurement theory\cite{Zurek}.Therefore the system is not subjected to any einselection rule and there will be no transfer of information from the system to the spin chain. This is consistent with the result of Yan et al. \cite{Yan et al.} if their parameters are set as $ \delta=0 \mbox{ and } c_{y}=-c_{x}=-c_{z}=1 $, namely by preparing their two-qubit system in $ |\psi\rangle=\frac{1}{2}(|01\rangle + |10\rangle)$. In addition, it is also consistent with the result of Guo et al. when the environment is prepared in the thermal state and QCs measured by the lower bound of concurrence and bipartite quantum discord \cite{Guo et al. 2013}.Therefore, in the spin chain induced decohence process, the form of pointer states are independant with respect to the preparation way of both the system and the spin chain, what is consistent with our intuition. Futhermore, from the symetric nature among those pointer states it comes that, the general form of the pointer states are parametric defined (in terms of number of qubits subjected to the mesurement) by the environmental spin chain. The parametric pointer state associated to the work of Yan et al. (namely, $ |\psi\rangle $ mentioned above) and the present work is $ |\psi\rangle = \frac{1}{\sqrt{N}}\sum_{i=0}^{N-1}|2^{i}\rangle $. Another one can be obtained from the previous one by complementing the binary expressin of each basic state, namely, $ |\psi'\rangle = \frac{1}{\sqrt{N}}\sum_{i=0}^{N-1}|2^{N}-1-2^{i}\rangle $. In this way, a new perpective appears to the conservation of information in quantum information processing. Indeed, one can conserve the information in an arbitrary large multipartite system, by preparing it into its associate state with respect to a generic pointer state of the surrounding.
	\\
	
	\begin{figure}[!htb]
		\includegraphics[scale=0.27]{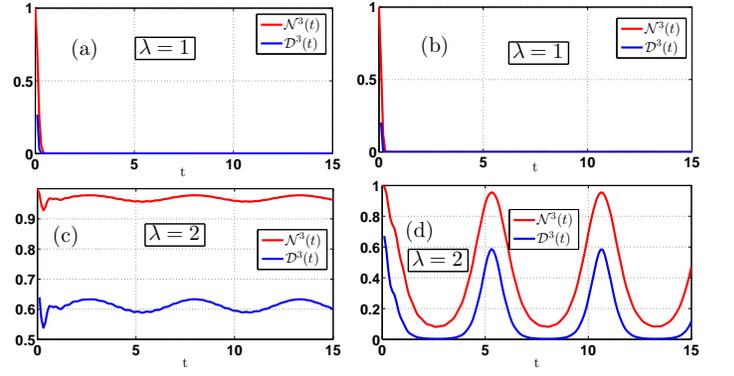}
		\caption{QCs against time for different magnetics fields . (a) and (c) are plotted for environment prepared in the ground state ; (b) and (d) in the vacuum state. D=0.5, N =400 , g=0.05, $ \gamma=0.4 $}
		\label{f7}
	\end{figure}
	\begin{figure}[!htb]
		\includegraphics[scale=0.27]{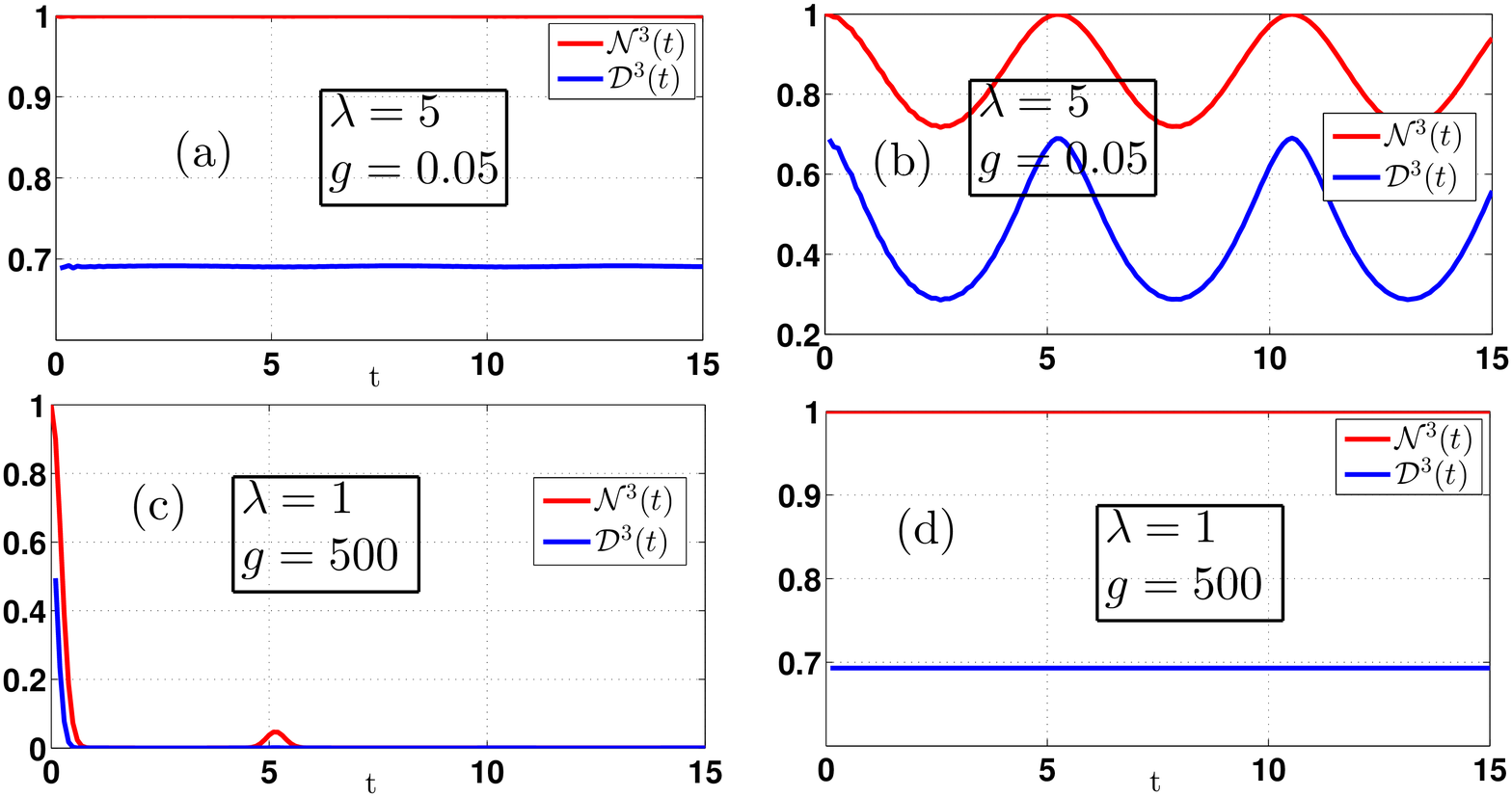}
		\caption{QCs against time for different magnetics fields . (a) and (c) are plotted for environment prepared in the ground state ; (b) and (d) in the vacuum state. D=0.5, N =400, $ \gamma=0.4 $}
		\label{f8}
	\end{figure}
	
	When the system is prepared in the GHZ state, namely, when $a=1$ the QCs read as:
	\begin{equation}
\left\lbrace 
\begin{split}
&\mathcal{N}^{(3)}(\rho_{ABC}(t))=|F_{07}(t)|\\&
\mathcal{D}^{(3)}(\rho_{ABC})=\frac{1+|F_{07}|}{2}\log(1+|F_{07}|)\\& + \frac{1-|F_{07}|}{2}\log\left( 1-|F_{07}|\right) 
\end{split}.
\right.
\end{equation}
	These expressions are similar to the ones obtained by Yan et al. \cite{Yan et al.} in the case of two-qubit initially prepared in $ |\psi\rangle=\frac{1}{2}(|00\rangle+|11\rangle) $ and the spin chain environment in the ground state. We obtain their result for uniformly system-environment coupling strength by substituting $g$ by $\frac{2}{3}g$. In addition it is similar to the one obtained in ref.\cite{Guo et al. 2013} when their spin chain environment is prepared in the thermal state. From the similitude among those results, it comes that, given an environment, at least for the spin chain environment, the form of the QCs of a system embedded inside, namely its qualitative behaviour, does not change when we set up the number of qubits if the preparation of the system is done in such a manner to conserve the symetry among its components.
	\\\hspace*{0.5cm}Let us note that, when 
	$
	|F_{07}(t)| \rightarrow 1$ then $\mathcal{D}^{(3)}(\rho_{ABC}(t)) \mbox{, } \mathcal{N}^{(3)}(\rho_{ABC}(t)) \rightarrow 1
	$ and when $ |F_{07}(t)| \rightarrow 0 $ then $ \mathcal{D}^{(3)}(\rho_{ABC}(t)) \mbox{, } \mathcal{N}^{(3)}(\rho_{ABC}(t)) \rightarrow 0
	$.
	Therefore, from the heuristic analysis of section 3, it comes that, the QCs will vanish in very short time under many conditions as indicated by the numerical results of Fig.~\eqref{f7}(a) and (b). But the QCs can be robust in presence of strong magnetic field, Fig.~\eqref{f7}(c),(d) and conserved in the case of evolution from vacuum state strongly coupled to the system Fig.~\eqref{f8}(d) or by choosing the spin chain to be the XX model. Fig.~\eqref{f8} shows that the dynamic of QCs when the spin chain evolving from the ground state in presence of very strong magnetic field are equivalent to its dynamic when the spin chain evolves from the vacuum state strongly coupled to the central spins. From analytical and numerical results it comes that Tripatite Negativity is more robust that GTQD.
	
	\section{Conclusion}\label{s5}
	We have studied the decoherence process induced by a spin chain environment on a central spins consisting of R spins and we have applied the results on the dynamics of QCs of three interacting qubits. In order to see the impact of the initial prepared state of the spin chain environment on decoherence process, we have assumed the spin chain environment prepared in the two main ways, namely, either the ground state or the vacuum state in the momentum space. In addition, in order to investigate the effect of initial central spins prepared state on the survival of QCs, we have made a parametric computation of tripartite negativity and GTQD for the central spins prepared in either the GHZ or the W state which are the well known (experimentally and theoretically) tripartite entangled states.
	\\ \hspace*{0.5cm}About decoherence process, we have shown that, in the weak coupling regime at the quantum critical point, the decoherence process appear exponentially in a short time limit, with the second power of time for evolution from ground state and four power for evolution from vacuum state. In addition, in the case of evolution from  the vacuum state strongly coupled to the central system, the spin chain environment will not induce any decoherence process while it behaves oscillatory or as an oscillatory Gaussian envelop for the spin chain prepared in the ground state. In the weak coupling regime with strong magnetic field the evolution from ground state induces a slight decoherence so that the information content by the central spins system will be more robust far from the quantum critical point than at the quantum critical point where the decoherence appeared exponentially.
	\\ \hspace*{0.5cm}About QCs, we have shown that, in the case of evolution from W state QCs remain constant independently to the environment prepared state. We have related this to the einselection rule introduced by Zurek. However, in the case of evolution from GHZ state, QCs obey to same dynamics as one of the decoherence factors and thus vanish in short time limit in many conditions.
	\\ \hspace*{0.5cm}From the decoherence behavior investigated in section 3 and the expressions of QCs obtained in section 4, it comes that the degradation of QCs due to decoherence can be reverse in presence of strong magnetic field, in the case of evolution from vacuum state strongly coupled to the system, by choosing the spin chain to be the XX model or by preparing the system in the W state.

%\appendix
%\section{}

%\begin{acknowledgments}
%\end{acknowledgments}


\bibliography{basename of .bib file}

\begin{thebibliography}{50}
	\bibitem{Zurek} Wojciech Hubert Zurek, Rev. Mod. Phys. \textbf{75}, 715 (2003).
	\bibitem{Horedecki et al.} R. Horodecki, P. Horodecki, M. Horodecki, and K. Horodecki, Rev. Mod. Phys. \textbf{81}, 865 (2009).
	\bibitem{Ollivier et al.} H. Ollivier and W. W. Zurek, Phys. Rev. Lett. \textbf{88}, 017901 (2001); L. Henderson and V. Vedral, J. Phys. A \textbf{34}, 6899 (2001); S. Luo, Phys. Rev. A \textbf{77}, 042303 (2008).
	\bibitem{Yan et al.} Y.-Y Yan, L.-G. Qin, and L.-J Tian, Chinese Physics B \textbf{21}, 100304 (2012).
	\bibitem{Guo et al. 2013} J.L. Guo, and G.-L. Long, Eur. Phys. J. D \textbf{67}, 53 (2013).
	\bibitem{Yang et al.} Y. Yang, and A.-M. Wang, Chin. Phys. B \textbf{23}, 020307 (2014).
	\bibitem{You et al.} W.-L. You and Y.-L. Dong, Eur. Phys. J. D, textbf{57}, 439 (2010).
	\bibitem{Hu} M.-L Hu, Phys. Lett. A, \textbf{374}, 3520 (2010).
	\bibitem{Xie et al.} L.-J Xie, D.-Y Zhang, and X.-Wn Wang, Theor. Phys. \textbf{50}, 3096 (2011).
	\bibitem{Guo et al. 2009} Guo J.-L, Jin J.-S and Song H.-S, Physica A \textbf{388}, 3666 (2009).
	\bibitem{Guo et al. 2011} J.L. Guo, and H.S. Song, Eur. Phys. J. D \textbf{61}, 791 (2011).
	\bibitem{Ali et al} M. Ali, A. R. P. Rau, and G. Alber, Phys. Rev. A 81, 042105 (2010).
	\bibitem{Quan et al.} H.T. Quan, Z. Song, X.F. Lieu, P. Zanardi and C.P. Sun, Phys. Rev. Lett. \textbf{96}, 140604 (2006).
	\bibitem{Yuan et al.} Z.-G Yuan, P. Zhang, and S.-S Li, Phys. Rev. A \textbf{76}, 042118 (2007).	
	\bibitem{Kenfack et al.} L. T. Kenfack, M. Tchoffo, G. C. Fouokeng, L. C. Fai, Int. J. Quantum Inf. \textbf{15}, 1750038 (2017).
	\bibitem{Cheng et al.} Cheng W. W, Liu J.-M, Phys. Rev. A \textbf{79}, 052320  (2009).
	\bibitem{Jafarpour et al.} M. Jafarpour, F. K. Hasanvand and D. Afshar, Commun. theor. Phys. \textbf{67}, 27 (2017).
	\bibitem{Sun et al.} Z. Sun, X. Wang and C. P. Sun, Phys. Rev. A \textbf{75}, 062312 (2007).
	\bibitem{Cucchietti} F. M. Cucchietti, J.P. Paz, and W. H. Zurek, Phys. Rev. A \textbf{72}, 052113 (2005); F. M. Cucchietti, S. F.-Vidal, and J. P. Paz, Phys. Rev. A \textbf{75}, 032337 (2007).
	\bibitem{Beggi et al} A. Beggi, F. Buscemi, and P. Bordone, Quantum Inf. Process. \textbf{14}, 573 (2015).
	\bibitem{Jafarpour et al.} M. Jafarpour, F. K. Hasanvand, and D. Afshar, Commun. Theor. Phys. \textbf{67}, 27 (2017).
	\bibitem{Tian et al.} L.-J Tian, Y.-Y Yan, and L.-G Qin, Commun. Theor. Phys. \textbf{58}, 39 (2012).
	
	
	
\end{thebibliography}

\end{document}